\documentclass{article}
\usepackage{authblk}
\usepackage[utf8]{inputenc}
\usepackage{indentfirst}
\usepackage[T1]{fontenc}
\usepackage{multirow}
\usepackage{rotating}
\usepackage{vmargin}
\usepackage{amsfonts}
\usepackage{cite}
\usepackage{url}

\title{Propensity Score Matching underestimates Real Treatment Effect, in a simulated theoretical multivariate model}

\author[1,2]{Daniel García Iglesias}
\affil[1]{Arrhythmya Unit. Cardiology Department. Hospital Universitario Central de Asturias. Oviedo. Spain.}
\affil[2]{Instituto de Investigación Sanitaria del Principado de Asturias. Oviedo. Spain}

\begin{document}

\maketitle

\begin{abstract}
Propensity Score Matching (PSM) is an useful method to reduce the impact ofTreatment - Selection Bias in the estimation of causal effects in observational studies. After matching, the PSM significantly reduces the sample under investigation, which may lead to other possible biases. In this sense, we want to analyse the behaviour of this PSM compared with other widely used method to deal with non-comparable groups, as is the Multivariate Regression Model (MRM). Monte Carlo Simulations are made to construct groups with different effects in order to compare the behaviour of PSM and MRM estimating this effects. Also the Treatment - Selection Bias reduction for the PSM is calculated. With the PSM a reduction in the Treatment - Selection Bias is achieved, with a reduction in the Relative Real Treatment Effect Estimation Error, but despite of this bias reduction and estimation error reduction, the MRM significantly reduces more this estimation error compared with the PSM. Also the PSM leads to a not insignificant reduction of the sample. This loss of information derived from the matching process may lead to another not known bias, and thus, to the inaccurate of the effect estimation compared with the MRM. 
\end{abstract}


\noindent \textbf{Abreviations:} PS: Propensity Score, PSM: Propensity Score Matching, RTE: Real Treatment Effect, GLM: General Linear Model, MRM: Multivariate Regression Model. 

\section{Introduction}

Propensity Score Matching (PSM) is an useful method to reduce the impact of treatment-selection bias in the estimation of causal effects in observational studies. Since firstly described by Rosenbaum and Rubin in 1983 \cite{Rosembaum1983}, its utility in Medicine, Psicology, Economics and other fields, has increased exponentially in the last years \cite{Austin2008}. Although it does not bypass the necessity for randomised studies, it may be an alternative to reduce the impact of treatment-selection bias in observational studies. 

The Propensity Score (PS) is defined as the subject’s probability of receiving a specific treatment conditional on the observed covariates\cite{Rosembaum1983}. After stratification by its PS, treated and untreated patients are matched by their PS with the most similar individuals of the opposite group. It leads to a more similar distribution of baseline characteristics between treated and untreated subjects, and it have been demonstrated that this method reduces the Treatment-selection bias\cite{Li2013,Innocenti2018}. 

Once two comparable groups have been obtained, researchers treat this studies like more similar to randomised studies (although it does not substitute this randomised studies), and use them as a reasonable alternative for observational studies\cite{Austin2009}. In this sense, it is thought that because PSM controls the possible treatment-selection bias, it would be possible to directly measure the effects in both matched groups, and thus, it may be better for observational studies than other multivariate adjustment methods. 

One important concern is the influence of the not insignificant loss of non matched individuals that may be seen in some works using this PSM method\cite{Seng2018}, and who are not used for posterior analysis. Because we need to eliminate enough unmatched individuals to guarantee in some way the Treatment - Selection Bias correction, it is not possible to know if the elimination of this unmatched individuals can cause some loss of information that in other ways would be analysed, and therefore lead us to a non controlled bias. In this sense other authors have reported the over-employment of this technique and its potential implications in potential biases\cite{King2018, MCMURRY2015}.

Moreover, although we are controlling the Treatment - Selection Bias, after matching we are directly measuring the effect in both matched groups. In this way, its behaviour compared with a multivariate adjustment method, which is widely used to control groups for other possible confounders, has never been tested. Given that there is a loss of individuals prior to the analysis, we do not know if this loss of information can variate the results obtained with the multivariate analysis. 

Because of that, we wanted to test the behaviour of PSM in different situations, compared with a multivariate analysis based on General Linear Models (GLM), in the estimation of treatment effects. For this purpose we developed a theoretical Montecarlo Method of treatment effects in which we applied the PSM and a multivariate analysis based on a GLM to compare their ability to estimate the Real Treatment Effect (RTE) in each situation. 

\section{Methods}

\subsection{Theoretical Multivariate model}

Suppose $Y_{z}(x)$ the patient $z$ probability for a certain event. Its probability may be influenced by a series of independent variables $A_{zj}$ and $B_{zk}$ each one of them with a concrete weight in this patient probability prediction $q_j$ and $s_k$ respectively. $A_{zj}$ variables may be related to the received treatment and $B_{zk}$ variables are supposed to be independent of the received treatment. It may be also influenced by the treatment status $X_{z}$ of the patient, which may confer some protection $t$ against the event under study. 

For each patient there may be also some unmeasured influence $\epsilon_{z}$ in its event probability, which may vary from patient to patient and may be because some unknown or non measured variables. We consider it as a random variable, which may be based in a normal distribution:
\begin{equation}
    \epsilon_{z} \sim \mathcal{N} (0,1) 
\end{equation}

Each patient may present some different characteristics. A part from this $A_{zj}$ and $B_{zk}$  characteristics that may predispose in some way the probability for the event under study, it may present other variables unrelated with the event under study. Some of them may predispose to receive the treatment under study $C_l$ and others $D_m$ may be unrelated to either patient outcomes or treatment predisposition. 

For a concrete patient the logit of the probability for a certain event may be predicted by the formula:
\begin{equation}
Y_{z}(x)= X_{z}t + A_{z1}q_{1} + ... + A_{zj}q_{j} + B_{z1}s_{1} + ... + B_{zk}s_{k} + \epsilon_{z}
\end{equation}

And thus the theoretical event probability for a $z$ given patient will be:
\begin{equation}
\rho_{z}= \frac {e^{Y_{z}(x)}} {1+e^{Y_{z}(x)}}
\end{equation}

\subsection{Monte Carlo Simulations}

Once the theoretical model is build, Monte Carlo Simulations are made to construct groups for a posterior analysis. Simulated experiments under this conditions are made, and an event status $\mathbb{E}_{z}$ is assigned to each patient $z$ in each simulated experiment, based on a binomial distribution with $\rho_{z}$ probability. 
\begin{equation}
\mathbb{E}_{z} \sim \mathcal{B}(n,\rho) = \mathcal{B}(1,\rho_{z})
\end{equation}

The theoretical (real) treatment effect is modified after each group of simulations, ranging from a $0$ (null) effect to a $5$ fold event reduction, in $0.1$ intervals. 

\subsection{Unadjusted Model}

The Odds Ratio for the event prevention under Treatment status is calculated with a univariable General Linear Model (GLM). For each experiment, with the complete matrix of events $Y$ and treatment $X$ status, a GLM is build to estimate the unadjusted estimated risk prevention effect for the treatment $t_{ua}$:
\begin{equation}
Y = Xt_{ua} + U
\end{equation}

From this built model, the estimated Odds Ratio for the RTE $OR_{ua}$ is calculated:
\begin{equation}
    OR_{ua} = e ^ {t_{ua}}
\end{equation}

Since this measured Odds Ratio does not take in account other variables, it will be named the unadjusted Odds Ratio and it will be considered the reference for the improvement in the RTE estimation.

\subsection{Multivariate Regression Model}

For control purposes a Multivariate Regression Model (MRM) is build for each experiment with all the variables under analysis. In each experiment the complete matrix of events $Y$, treatment status $X$ and analysed variables $A$, $B$, $C$ and $D$ are used to build the MRM and estimate the multivariate adjusted event reduction of the treatment $t_{multi}$: 
\begin{equation}
Y = Xt_{multi} + Aq_{multi} + Bs_{multi} + Cu_{multi} + Dv_{multi} + U
\end{equation}

From this built model, the estimated Odds Ratio for the RTE (Multivariate Odds Ratio, $OR_{multi}$) is calculated:
\begin{equation}
    OR_{multi} = e ^ {t_{multi}}
\end{equation}

It will be used as the gold standar for the  estimation of the RTE.

\subsection{Propensity Score Matching Model}

PSM is used to estimate the Treatment Effect. As described early, the PS is build based on a MRM designed to estimate each patient's predisposition to receive the treatment under investigation. In each patient the PS is calculated for posterior matching between treated and untreated individuals. For the matching the nearest method is used, with a caliper of 0.2.

Once matched is done, the treatment effect is estimated based on a GLM. For each experiment, the matrix of matched individuals with the events $Y_{match}$ and treatment $X_{match}$ status are used to build a GLM to estimate the estimated risk prevention effect for the treatment $t_{match}$:
\begin{equation}
Y_{match} = X_{match}t_{match} + U
\end{equation}

From this built model, the estimated Odds Ratio for the RTE $OR_{match}$ is calculated:
\begin{equation}
    OR_{match} = e ^ {t_{match}}
\end{equation}

\subsection{Statistical Analysis}

For each treatment effect situation, RTE Estimation is measured with each one of the three described methods. This RTE estimation is lately compared with the RTE to calculate the inaccurate of the RTE Estimation (Relative RTE Estimation Error). Al variables are expressed as mean +/- 95\% Confidence Interval. The comparison of the RTE Estimation Error between the three described methods is done with a multivariate analysis of variance (MANOVA test). 

For the MRM and the PSM Model, the RTE Estimation Error is compared with the Unadjusted Model, to calculate the Relative RTEE Estimation Error Reduction for each one. The comparison of this Relative RTEE Estimation Error Reduction between MRM and PSM Model is done with a paired T Test. 

For the Unadjusted Model and the PSM Model the Treatment-Selection Bias is calculated with a Pearson's chi-squared test to evaluate the homogeneity between groups. The comparison of the Treatment - Selection bias between both models is done with a paired T Test comparing the the chi-squared test statistic of both models, and the reduction in the Treatment - Selection bias with the PSM Model is expressed as mean +/- 95\% Confidence Interval.

For the PSM Model the percentage of excluded patients in each analysis will be also analysed. This will be expressed as the mean of the percentage exclusions in each analysis and the 95\% Confidence Interval of the percentage exclusions in each analysis. 

\subsection{Analysis Software}

For the data generation with the Monte Carlo Simulations and the posterior analysis, the open software R \cite{R} and the R library MatchIt \cite{MatchIt} were used. All wrote code for this purpose is available through the PropensityScoreReview repository \cite{Github}.

\section{Results}

\subsection{Monte Carlo Simulations}

A total of $5 \cdot 10^6$ Montecarlo Simulations are done. There are 50 blocks of experiments, each one with a different real treatment effect. In each block of experiments, each experiment is simulated with 500 individuals and later repeated 200 times for each real treatment effect. 

\begin{figure}[bt]
\centering
\includegraphics[width=14cm]{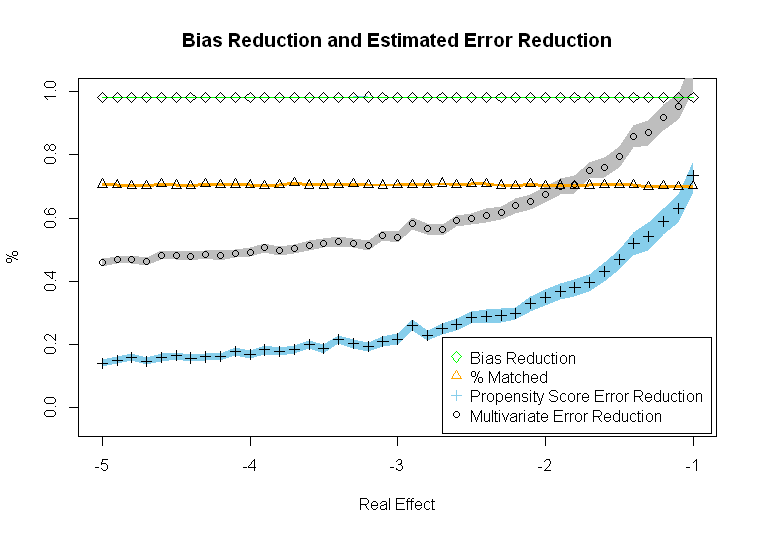}
\caption{Comparison of the Propensity Score Matching Model Treatment-Selection Bias reduction, Percentage of Matched Individuals and Relative Real Treatment Effect Estimation Reduction with the Multivariate Model Relative Real Treatment Effect Estimation Reduction.}
\end{figure}

\begin{table}
\centering
\begin{tabular}{|p{2.3cm}|p{2.3cm}|p{2.3cm}|p{1cm}|p{2.3cm}|p{2.3cm}|}
\hline
& \multicolumn{3}{c|}{T-S Bias} & &\\ 
\cline{2-4} 
Real Treatment Effect & Unadjusted Model & Matched Model & p Value & T-S Bias Reduction & Excluded Patients (\%) \\
\hline
5&220.7 (217.82, 223.57)&3.631 (3.479, 3.783)&<0.001&0.983 (0.983, 0.984)&70.66 (70.2, 71.12)\\
\hline
4&219.09 (216.36, 221.83)&3.599 (3.446, 3.751)&<0.001&0.983 (0.983, 0.984)&70.43 (69.98, 70.88\\
\hline
3&219.1 (216.29, 221.90)&3.698 (3.536, 3.861)&<0.001&0.983 (0.982, 0.984)&70.55 (70.1, 71)\\
\hline
2&217.6 (214.80, 220.33)&3.721 (3.576, 3.866)&<0.001&0.983 (0.982, 0.983)&70.12 (69.68, 70.56)\\
\hline
1&216.98 (214.31, 219.66)&3.738 (3.603, 3.873)&<0.001&0.982 (0.982, 0.983)&70.19 (69.72, 70.65)\\
\hline
0&219.47 (216.85, 222.09)&3.684 (3.532, 3.835)&<0.001&0.983 (0.982, 0.984)&70.21 (69.77, 70.65)\\
\hline
\end{tabular}
\caption{Treatment - Selection Bias for the Unadjusted Model and the Propensity Score Matching Model. Treatment - Selection Bias Reduction for the Propensity Score Matching Model. Percentage of excluded patients for the Propensity Score Matching Model. T-S Bias: Treatment - Selection Bias.}
\end{table}

\subsection{Treatment - Selection Bias reduction with Propensity Score Matching}

The PSM Model significantly reduces the Treatment - Selection Bias in all scenarios. As seen in table 1 and Figure 2, there is a Relative Treatment - Selection Bias reduction of about 0.98 in all scenarios. The main problem of this model is the important number of excluded patients (there is only a 70\% of patients that are included for the analysis). 

\subsection{Real Treatment Effect Estimation and Relative Real Treatment Effect Estimation Error Reduction with Propensity Score Matching Model and Multivariate Regression Model}

As it can be seen in figure 2 and 3, and table 2, the PSM Model and the MRM significantly estimate a more accurate RTE than the Unadjusted Model. This PSM Model and RMM present a significantly reduced Relative RTE Estimated Error, compared with the Unadjusted Model. 

The MRM also present a Relative RTE Estimated Error significantly lower than the PSM Model, which leads to a significantly increased Relative RTE Estimated Error Reduction compared with the PSM Model. 

\begin{figure}[bt]
\centering
\includegraphics[width=14cm]{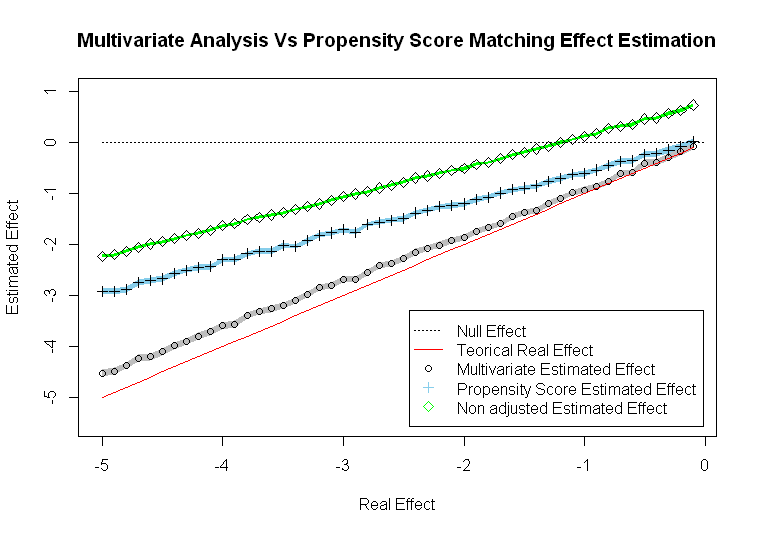}
\caption{Comparison of Real Treatment Effect and Real Treatment Effect Estimation from the Unadjusted Model, Multivariate Regression Model and Propensity Score Matching Model.}
\end{figure}

\begin{figure}[bt]
\centering
\includegraphics[width=14cm]{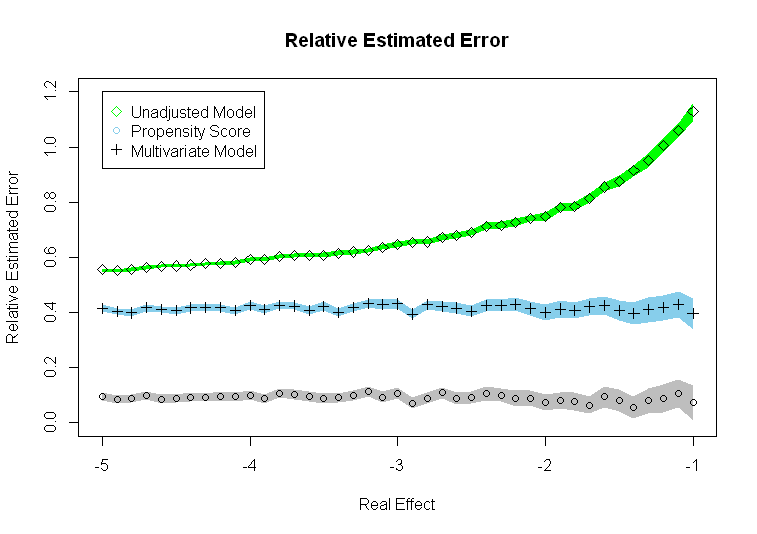}
\caption{Comparison of Relative Real Treatment Effect Estimation Error with the Unadjusted Model, Multivariate Model and Propensity Score Matching Model.}
\end{figure}

\begin{sidewaystable}
\centering
\begin{tabular}{|p{1.1cm}|p{1.8cm}|p{1.8cm}|p{1.8cm}|p{1cm}|p{1.8cm}|p{1.8cm}|p{1.8cm}|p{1cm}|p{1.8cm}|p{1.8cm}|p{1cm}|}
\hline
& \multicolumn{4}{c|}{Estimated Treatment Effect} & \multicolumn{4}{c|}{Relative RTE Estimated Error} & \multicolumn{3}{c|}{RTE Estimated Error Reduction}\\ 
\cline{2-12} 
Real Treatment Effect & Unadjusted Model & Matched Model & Multivariate Model & p Value & Unadjusted Model & Matched Model & Multivariate Model & p Value & Matched Model & Multivariate Model & p Value\\
\hline
5&-2.225 (-2.254, -2.195)&-2.922 (-2.984, -2.86)&-4.521 (-4.594, -4.448)&<0.001&0.555 (0.549, 0.561)&0.416 (0.403, 0.428)&0.096 (0.082, 0.11)&<0.001&0.139 (0.128, 0.151)&0.459 (0.446, 0.473)&<0.001\\
\hline
4&-1.63 (-1.659, -1.601)&-2.3 (-2.355, -2.245)&-3.597 (-3.659, -3.536)&<0.001&0.592 (0.585, 0.6)&0.425 (0.411, 0.439)&0.101 (0.085, 0.116)&<0.001&0.167 (0.154, 0.181)&0.492 (0.478, 0.506)&<0.001\\
\hline
3&-1.059 (-1.088, -1.030)&-1.707 (-1.759, -1.655)&-2.676 (-2.736, -2.616)&<0.001&0.647 (0.637, 0.657)&0.431 (0.41, 0.448)&0.108 (0.088, 0.128)&<0.001&0.216 (0.2, 0.232)&0.539 (0.522, 0.556)&<0.001\\
\hline
2&-0.504 (-0.536, -0.472)&-1.201 (-1.257, -1.144)&-1.852 (-1.91, -1.793)&<0.001&0.748 (0.732, 0.76)&0.4 (0.371, 0.428)&0.074 (0.045, 0.103)&<0.001&0.348 (0.324, 0.373)&0.674 (0.649, 0.698)&<0.001\\
\hline
1&0.129 (-0.096, 0.161)&-0.604 (-0.661, -0.548)&-0.927 (-0.99, -0.864)&<0.001&1.129 (1.096, 1.161)&0.396 (0.339, 0.452)&0.073 (0.01, 0.136)&<0.001&0.733 (0.684, 0.783)&1.056 (1.003, 1.109)&<0.001\\
\hline
0&0.743 (-0.708, 0.777)&0.026 (-0.032, 0.085)&-0.078 (-0.139, -0.017)&<0.001&8.426 (8.08, 8.772)&1.264 (0.676, 1.853)&0.22 (-0.394, 0.83)&<0.001&7.162 (6.622, 7.701)&8.206 (7.699, 8.714)&<0.001\\
\hline
\end{tabular}
\caption{Real Treatment Effect Estimation for the Unadjusted Multivariate and Propensity Score Matching models. Relative Real Treatment Effect Estimation Error for the Unadjusted Multivariate and Propensity Score Matching models. Relative Real Treatment Effect Estimation Error Reduction for the Multivariate and Propensity Score Matching models. RTE: Real Treatment Effect Estimation. T-S Bias: Treatment - Selection Bias.}
\end{sidewaystable}

\section{Discussion}

PSM has been widely used in different subjects for Treatment - Selection Bias reduction. As we shown in this present work, and as it was seen in previous works\cite{Austin2009}, with this method we can correct the Treatment - Selection Bias properly, obtaining two comparable groups, so we can thus directly measure the effect under investigation. In this sense, in our work the Treatment - Selection Bias practically disappears with the PSM (Treatment - Selection Bias reduction of 0.982 - 0.983 among all scenarios). 

As we mentioned earlier, our main concern about the PSM method is the percentage of unmatched individuals and its possible influence in the posetrior estimation of RTE. In our work results there is a not insignificant reduction of the sample, with a percentage of analysed individuals of 70.12 to 70.66\% from the total of individuals under investigation. And what we can not know is if this important reduction in the population for analysis may lead to another non controlled bias, since the excluded patients may have some characteristics that we are obviating for posterior analysis. 

Other widely used methods, such as MRM, do not deal with two comparable populations, but instead, they weigh the different variables under study. 
Despite dealing with non-comparable groups, thanks to this weighting of the analysed variables, it can solve in a different way the problem of the Treatment - Selection Bias. Also, since no individual is eliminated from the analysis, there is no loss of information, reducing other potential biases that may appear in the PSM method.

In our present work, although both (PSM and MRM) reduce the Relative RTE Estimation Error, this reduction is better with the MRM, compared with the PSM. This best performance of the MRM may confirm our previous preoccupation about the possible influence of the sample reduction on the posterior estimation of the RTE. 

Since there may be a significant reduction of the sample under analysis when we are using a PSM method, we have to take it in account before accepting the obtained results, specially when this reduction is important. Other multivariate methods should always be done, in addition to the PSM analysis, and both results compared, in order to seek for a possible uncontrolled bias. If a significant difference is obtained between both analysis, we have to suspect a possible bias derived from the sample reduction, once the matching has been done.


\end{document}